%% file: main.tex
\documentclass[final,5p,times,twocolumn,numbers]{elsarticle}
\usepackage{graphicx}

\usepackage{amssymb}
\usepackage{minted}
\usepackage{lipsum}
\usepackage{xcolor}
\usepackage{tabularx}
\usepackage{booktabs}
\usepackage{listings}

\usepackage{mathtools}

\usepackage[normalem]{ulem}
\useunder{\uline}{\ul}{}
\usepackage{booktabs}

\usepackage{bbding}
\usepackage{pifont}
\usepackage{amssymb}
\usepackage{utfsym}

\usepackage{multirow}

\usepackage{wasysym}

\usepackage{amsmath,amssymb}

\newcommand{\halfcheck}{\checkmark\kern-1.1ex\raisebox{.7ex}{\rotatebox[origin=c]{125}{--}}}

\usepackage[colorlinks]{hyperref}
\hypersetup{
  colorlinks   = true,
  citecolor    = blue,
  linkcolor    = black,
  urlcolor     = blue,
  filecolor    = magenta
}

\journal{Journal of Network and Computer Applications}
\biboptions{sort&compress}

\begin{document}

\begin{frontmatter}



\title{AgenticOS: An Intent-Oriented Secure Operating System Architecture for Autonomous AI Agents}



            

\author[4]{Zhen Zhao}
\ead{jeremiazhao@tencent.com}

\author[1]{Yu Zhang}
\ead{yugozhang@tencentos.com}

\author[3]{Yanpeng Zhu}  
\ead{rockerzhu@tencent.com}

\author[2]{Jia Wang}  
\ead{jasperwang@tencent.com}

\author[2]{Songqiao Tao}  
\ead{toeytao@tencent.com}

\author[2]{Xin Cheng}  
\ead{denisecheng@tencent.com}

\author[5]{Jiexin Gao \corref{cor1}}  
\ead{gjx@mail.scuec.edu.cn}


\address[1]{Tencent Technology (Shanghai) Co., Ltd., Shanghai 200030, China}
\address[2]{Shenzhen Tencent Computer System Co., Ltd., Shenzhen 518057, China}
\address[3]{Tencent Cloud Computing (Beijing) Co., Ltd. Shanghai Branch, Shanghai 200030, China}
\address[4]{Tencent Cloud Computing (Beijing) Co., Ltd., Beijing 100080, China}
\address[5]{Office of Informatization Development and Management, South-Central Minzu University, Wuhan 430074, China}


\input{chapter/abstract}

\end{frontmatter}


\input{chapter/intro}

\input{chapter/design}

\input{chapter/agenticos}

\input{chapter/intent}
\input{chapter/security}

\input{chapter/application}

\input{chapter/discussion}

\input{chapter/conclusion}


\bibliographystyle{elsarticle-num-names}
\bibliography{chapter/ref} 

\end{document}

%% file: chapter/abstract.tex
\begin{abstract}
As LLM-driven autonomous agents gradually acquire capabilities for autonomous planning, tool invocation, network access, code execution, and cross-application collaboration, traditional operating-system security models based on “resource exposure plus permission checks” are facing structural challenges. Existing POSIX-style system-call interfaces expose general-purpose resource primitives -- such as files, networks, processes, memory mappings, and dynamic execution -- to processes. Once an agent runtime is compromised by prompt injection, supply-chain poisoning, or malicious tool outputs, an attacker may combine these primitives into behaviors far beyond the scope of the user’s task authorization. This paper proposes AgenticOS, an intent-oriented secure operating system architecture for autonomous AI agents. Its core paradigm is to reconstruct the operating system from a “resource manager” into an “intent filter”: agents no longer request low-level resources directly, but instead submit structured intent declarations. Based on the Manifest, the system dynamically synthesizes a least-capability environment and enforces mandatory mediation, auditing, and information-flow constraints on external effects. AgenticOS does not attempt to weaken the general computational capability of the operating system. Rather, it redraws the ownership boundary of general capabilities: generic file, network, protocol, process, and toolchain capabilities are no longer directly exposed to the agent runtime, but are encapsulated on the system side as policy-constrained, auditable, and compositionally limited semantic capabilities. Building on prior work in capability security, microkernels, sandboxing, formal isolation, and information-flow control, this paper proposes a four-layer architecture consisting of the Ghost Kernel, Logic Shutter, Agent Capsule, and Semantic Boundary Gateway. It further introduces the Intent ABI, Manifest-Only Runtime, Weaver-based dynamic capability generation, and an admission model for AgenticOS-native Skills. The paper also discusses the ecological boundary of application capability migration: the long-term goal of AgenticOS is not to replace all application forms, but to gradually consolidate software capabilities that are delegable, amenable to semantic abstraction, and auditable into operating-system-native capabilities.

\end{abstract}

\begin{keyword}
Operating System Security \sep AI Agent \sep Intent Filtering \sep Capability Security \sep Intent ABI \sep Formal Isolation

\end{keyword}

%% file: chapter/intro.tex
\section{Introduction}
\label{introduction}

\subsection{System Security Problems Introduced by Autonomous Agents}
AI is undergoing a transition from text-generation models to autonomous agents capable of autonomous planning, tool invocation, and network interaction. Traditional applications are usually operated explicitly by users through graphical interfaces or command lines, and their behavioral boundaries are jointly determined by user actions, application logic, and operating-system permissions. By contrast, autonomous agents typically operate in the following mode: a user provides a high-level task, and the agent decomposes the task into steps, selects tools, reads context, accesses external services, and produces external effects on its own.

This shift creates a fundamental security tension: users authorize “task intent,” while operating systems expose “resource primitives.” For example, a user may only authorize an agent to “organize the experimental results of a project and generate a summary.” Yet in a traditional system, the agent process often simultaneously obtains general capabilities such as file reading, network access, subprocess execution, dynamic loading, and temporary-directory writes. If the agent is compromised by prompt injection, malicious dependencies, or poisoned tool outputs, an attacker can combine these general capabilities into lateral movement, sensitive-data exfiltration, persistence, or covert communication paths.

Therefore, the operating-system security problem in the agent era is not simply “how to grant an agent fewer permissions.” The deeper question is how the system can understand and constrain the forms of tasks that an agent is allowed to complete. This paper refers to this problem as the mismatch between resource-permission models and task intent.

\subsection{From Resource Manager to Intent Filter}
The POSIX system-call interface of current operating systems, such as Linux, is a set of general computational primitives. Interfaces such as $open$, $read$, $write$, $socket$, $connect$, $mmap$, $execve$, and $fork$ are neutral in isolation, but their compositional closure is extremely powerful: once a process obtains sufficient resource access, it can express behaviors far beyond the original task objective.

The central observation of AgenticOS is that, for highly autonomous agents, the security boundary should not be built around the low-level question of “whether a process is allowed to access a resource.” Instead, it should be built around the semantic question of “whether this external effect conforms to the declared task intent.” AgenticOS therefore aims to remove the agent runtime’s direct dependence on general system-call semantics and replace it with one-shot capability synthesis based on intent declarations. We position AgenticOS as an “agent-native operating system”: a secure computing substrate designed for highly autonomous AI systems, centered on intent constraints and least capability.

It should be noted that AgenticOS does not deprive the system of general computational capability. Instead, it encapsulates such capability on the system side as structured, auditable, and policy-mediated semantic capabilities, rather than handing it to agents in the form of raw system calls or general resource handles.

\subsection{Theoretical Foundations and Limitations of Existing Approaches}
AgenticOS is not a security model invented from scratch; rather, it recombines and lifts several existing lines of work.

First, capability security emphasizes least authority and unforgeable capability handles. Traditional capability-based systems, the object-capability model, KeyKOS, EROS, Capsicum, and CHERI all attempt to replace global permission checks with capability objects \cite{1, 2, 3, 4, 5, 6}. AgenticOS inherits this idea, but further elevates capability objects from “resource handles” to “task-intent handles”: a capability constrains not only which resource can be accessed, but also under what semantics and in which task context an external effect may be produced.

Second, microkernels and formal verification emphasize a minimal trusted computing base (TCB) and provable isolation. Systems such as seL4 demonstrate that kernel-level isolation and access control can be rigorously proven on a relatively small code base \cite{7}. The Ghost Kernel in AgenticOS follows this direction, but it does not attempt to become a complete general-purpose kernel. Instead, as the minimal trusted kernel, the Ghost Kernel provides only isolation, scheduling, measurement, and the root of attestation.

Third, sandboxing, WASM, eBPF, SFI/CFI, and language-level restrictions can constrain the execution environment of untrusted code \cite{8, 9, 10}. However, sandboxing alone cannot solve the problem of “legal capabilities being abused in combination.” For instance, an agent that is allowed to read an internal API, summarize it through a model, and write a report may still exfiltrate sensitive information through a chain of legitimate interfaces. AgenticOS therefore introduces intent mediation, information-flow labels, and capability-composition analysis beyond sandboxing; these ideas can also be connected to classic information-flow control models \cite{11, 12}.

Fourth, LLM agents face new threats such as prompt injection, indirect prompt injection, tool-output poisoning, supply-chain contamination, and data exfiltration \cite{13}. Meanwhile, works such as ReAct and Toolformer show that models are evolving from pure text generation toward an execution paradigm that combines reasoning, acting, and tool use \cite{14, 15}. This paper argues that, for long-running agents that can invoke multiple tools, the security boundary should be lowered into the operating-system capability model rather than relying only on prompt filtering, tool allowlists, or application-layer sandboxes.

\subsection{Contributions}
The main contributions of this paper are as follows:

\begin{itemize}
  \item It proposes an intent-oriented operating-system security paradigm for autonomous agents, reconstructing the OS from a “resource manager” into an “intent filter.”

  \item It designs a four-layer AgenticOS architecture consisting of the Ghost Kernel, Logic Shutter, Agent Capsule, and Semantic Boundary Gateway.

  \item It introduces the Intent ABI and Manifest-Only Runtime, enabling the agent runtime to access only semantic capabilities generated from task declarations.

  \item It presents a Weaver-based dynamic capability generation mechanism, as well as generation, registration, and admission principles for AgentOS-native Skills.

  \item It analyzes the architecture’s impact on system-call attack surfaces, supply-chain poisoning, capability-composition attacks, intent drift, and covert channels.

  \item It discusses the ecological boundary of migrating from applications to application capabilities, and explains how AgentOS-native Skills can host delegable software capabilities.

\end{itemize}

%% file: chapter/design.tex
\section{Design Goals and Threat Model}
\label{Design Goals and Threat Model}

\subsection{Threat Model}

This paper focuses on the following in-scope threats:

\begin{itemize}
  \item Fully compromised agent process: an attacker obtains arbitrary code execution inside an Agent Capsule and attempts lateral movement, escape from the isolation boundary, or kernel-level operations.

  \item Supply-chain poisoning: agent dependencies, Skills, model configurations, or toolchain components are implanted with malicious logic that attempts to leak sensitive data through legitimate ABIs.

  \item Tool-output poisoning: external services return malicious instructions or structured inducements that cause the agent to perform unauthorized behavior.

  \item Capability-composition attacks: the attacker does not violate any single interface policy, but combines multiple legitimate interfaces into unintended external effects.

  \item Covert-channel attacks: the attacker uses call frequency, output size, timing differences, or interface presence to establish low-bandwidth illicit communication channels.
  
\end{itemize}

\subsection{Security Assumptions}
The security analysis of AgenticOS relies on the following assumptions:

\begin{itemize}
  \item Effective isolation primitives: hardware-level memory isolation, virtualization isolation, or trusted execution environments are logically reliable, such as VT-x, AMD-V, Intel TDX, AMD SEV, or equivalent mechanisms.

  \item Trusted Ghost Kernel: the minimal trusted kernel, the Ghost Kernel, has a sufficiently small code base, and its key isolation, scheduling, and measurement logic can be formally verified or intensively audited.

  \item Reliable cryptographic mechanisms: the cryptographic primitives used by capability tokens, component signatures, remote attestation, and audit-log hash chains are not broken.

  \item Trusted policy inputs: Manifests, organizational policies, human approvals, and capability-registration information come from a trusted management plane.
\end{itemize}

This paper does not assume that the agent itself is trustworthy, nor does it assume that an LLM can reliably identify all malicious instructions. The goal of AgenticOS is precisely to ensure that a compromised or misled agent still cannot cross the semantic capability boundary granted by the system.

\subsection{Design Goals}
AgenticOS has the following design goals:

\begin{itemize}
  \item Intent first: the system authorizes by task intent and external effect, rather than by raw files, raw ports, or raw processes.

  \item Least capability: before an agent starts, it must submit a Manifest, and the system generates only the minimum capability set required to complete the declared task.

  \item No raw byte streams: an Agent Capsule does not directly access sockets, pipes, device files, or arbitrary protocol byte streams.

  \item No arbitrary execution: an Agent Capsule does not directly expose interfaces such as $exec$, $fork$, $dlopen$, or $mmap$, which can be generalized into arbitrary execution environments.

  \item Auditability: all external effects must be recorded as structured events and bound to the Manifest, capability tokens, and call chains.

  \item Provable isolation: the underlying isolation, measurement, and capability-generation chains should be formally modelable, with the TCB minimized as much as possible.

  \item Evolvability: the system allows new Skills to be developed and registered, but does not allow agents to bypass the admission process and self-expand capabilities at runtime.
\end{itemize}


%% file: chapter/agenticos.tex
\section{AgenticOS Design}

\subsection{From Resource Manager to Intent Filter}

\begin{figure}[H]
    \centering
    \includegraphics[width=0.7\linewidth]{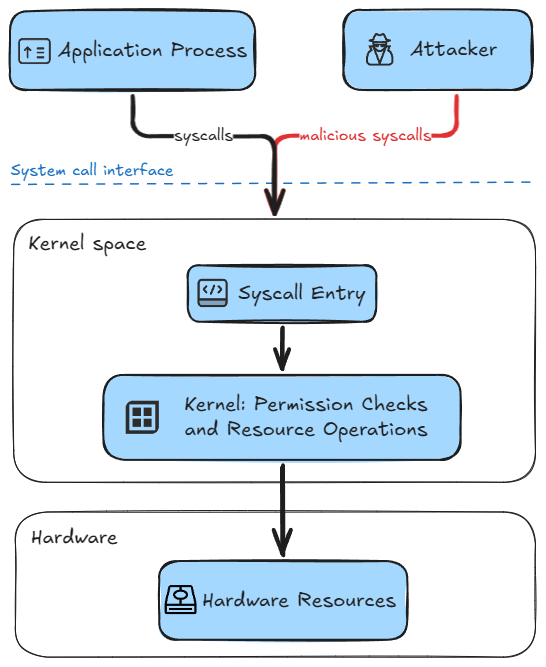}
    \caption{Traditional OS: Permission Based}
    \label{fig:traditional-os}
\end{figure}

\begin{figure}[H]
    \centering
    \includegraphics[width=0.7\linewidth]{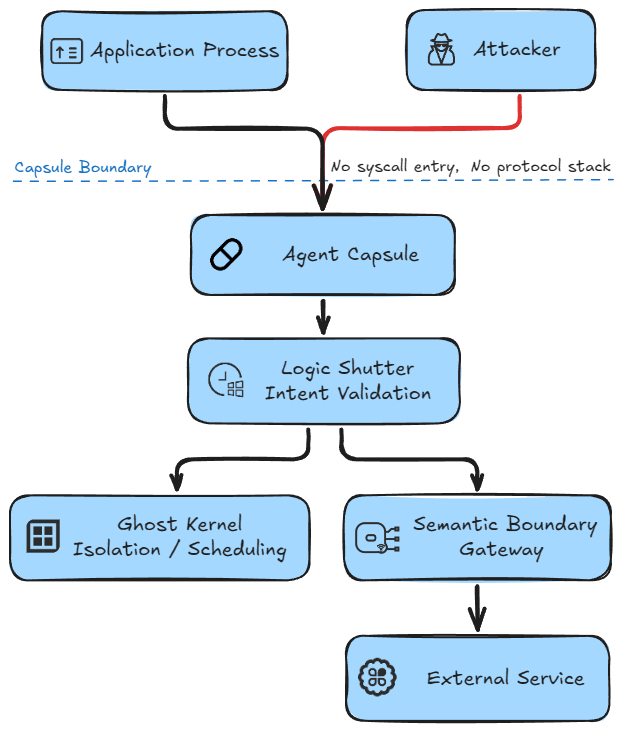}
    \caption{AgenticOS: Intent-Manifest Based}
    \label{fig:agentic-os}
\end{figure}

The core responsibility of a traditional OS is to manage hardware resources and check permissions when processes access those resources. AgenticOS shifts this responsibility toward intent filtering: the system first determines whether the external effect requested by an agent conforms to the task intent declared in the Manifest, and only then maps it to controlled low-level operations.( Fig \ref{fig:traditional-os} and Fig \ref{fig:agentic-os} illustrate the differences in security paradigms between traditional operating systems and AgenticOS.)

Under this model, the attack surface shifts from “finding exploitable system-call entries” to “attempting to abuse restricted semantic interfaces.” This does not mean that attacks become impossible; rather, it means that attackers must operate within a narrower, more auditable, and more policy-constrained abstraction layer.

\subsection{Four-Layer Architecture Overview}

AgenticOS adopts a four-layer vertical design, with strict one-way dependencies between layers to provide defense in depth.

The four layers have distinct responsibilities. The Ghost Kernel provides the minimal trusted isolation substrate. The Logic Shutter performs intent recognition, policy mediation, capability-token management, and auditing. The Agent Capsule hosts the restricted agent runtime. The Semantic Boundary Gateway handles external protocols, credentials, content filtering, and output normalization.

\subsection{Ghost Kernel: The Minimal Trusted Kernel}
The Ghost Kernel is the trust root of the entire system and runs at the highest privilege level, such as VMM Root, a hardware TEE secure domain, or an equivalent isolation domain. The term “Ghost Kernel” does not imply mystery; it emphasizes that the kernel is not directly reachable at runtime. It does not accept general system calls from Agent Capsules, expose device files, provide debugging interfaces, or host complex protocol stacks.

The Ghost Kernel is responsible for only three classes of primitives:
\begin{itemize}
    \item Encrypted memory allocation: based on hardware memory encryption or page-table isolation mechanisms, it allocates isolated physical pages for each Agent Capsule and enforces isolation between capsules through EPT/NPT or equivalent mechanisms.

    \item Deterministic time-slice scheduling: it provides controlled CPU-time allocation to reduce the ability to construct covert channels through scheduling timing.

    \item Measurement and attestation root: it provides cryptographic hashes of the capsule’s initial state, ABI Stubs, Manifest bindings, and semantic-library versions to support remote attestation and audit traceability.
\end{itemize}

\begin{figure}[H]
    \centering
    \includegraphics[width=0.9\linewidth]{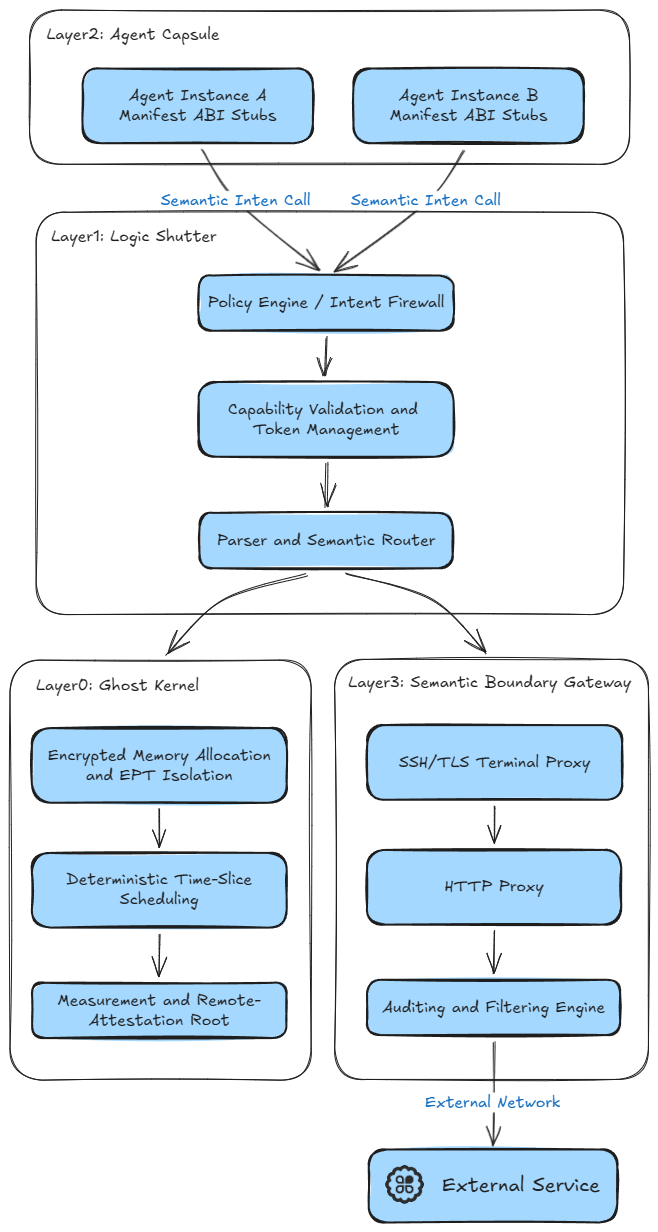}
    \caption{AgenticOS Four-Layer Architecture Diagram}
    \label{fig:Architecture-diagram}
\end{figure}

The core constraint of the Ghost Kernel is “no exposed interface.” Resource allocation is performed statically as much as possible when a capsule is created. At runtime, the Ghost Kernel responds only to a minimal control-plane request set from the Logic Shutter. Under the stated security assumptions, this design structurally removes privilege-escalation paths that rely on system-call entries and complex kernel services.

\subsection{Logic Shutter}
The Logic Shutter is the semantic translation layer and policy enforcement point of AgenticOS. It runs at a trust level above Agent Capsules and below the Ghost Kernel. It is responsible for intent recognition, capability validation, policy mediation, information-flow labeling, and audit logging.

The main functions of the Logic Shutter include:
\begin{itemize}
    \item Intent parsing and validation: it receives semantic requests from Agent Capsules and determines whether they match the capability list, data-flow constraints, and output targets declared in the Manifest.

    \item Capability-token management: it issues, validates, and revokes unforgeable capability tokens, binding them to agent identity, Manifest, request chain, and resource budget to prevent replay or cross-task reuse.

    \item Audit-log generation: it records each intent call as an immutable structured event, including caller, $cap\_id$, input digest, output digest, policy version, and mediation result.

    \item Information-flow labeling: it attaches source labels to input data and enforces label-based blocking, redaction, or human-confirmation policies at output interfaces.
\end{itemize}

The Logic Shutter is responsible for semantic recognition and intent mediation, whereas the Semantic Boundary Gateway is responsible only for executing external effects after the intent has been authorized. In this paper, the word “semantic” in “Semantic Boundary Gateway” does not refer to natural-language semantic recognition. Rather, it means that the gateway accepts only semantic operations structurally defined by the Intent ABI and does not expose raw byte streams, sockets, protocol stacks, or credentials to Agent Capsules.

A typical intent-call flow is shown below.

\begin{figure*}[h]
    \centering
    \includegraphics[width=1.0\linewidth]{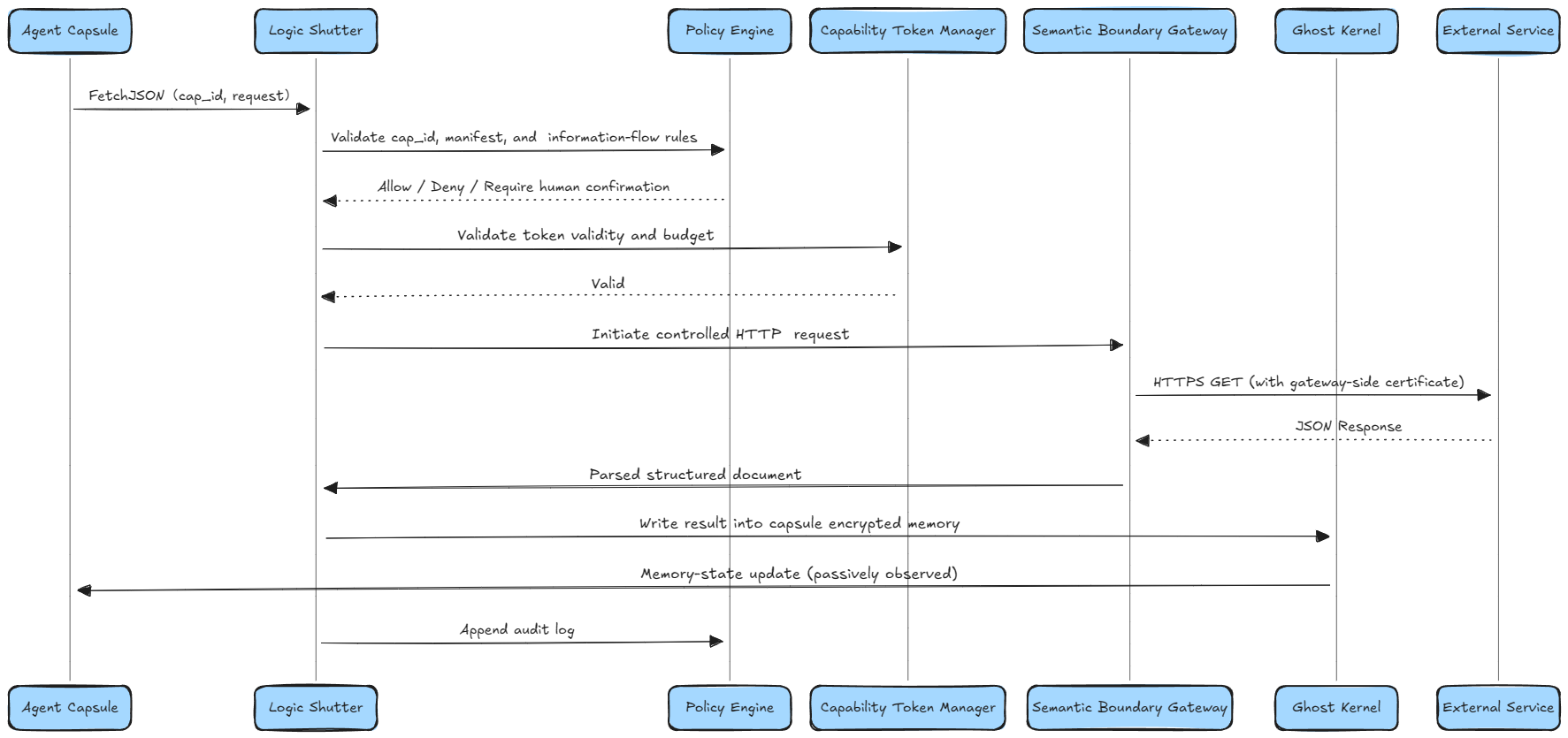}
    \caption{Intent invocation end-to-end flow—from capsule request to external response.}
    \label{fig:Sequence-diagram}
\end{figure*}

\subsection{Agent Capsule}
An Agent Capsule is the actual runtime environment for agent code. Unlike a traditional container or virtual machine, a capsule follows the Manifest-Only Runtime principle: before startup, the agent must submit a structured intent declaration that specifies required network domains, file scopes, tool-call types, model sessions, output targets, and human-confirmation points. Capabilities outside the Manifest have no corresponding interfaces in the capsule address space.

The key constraints of a capsule include:
\begin{itemize}
    \item Intent manifest discipline: all capabilities originate from the Manifest and are explicitly bound to the call chain through $cap\_id$.

    \item Dynamic ABI generation: at startup, the Logic Shutter and Weaver generate a dedicated interface table for the capsule based on the Manifest. If network access is not declared, the capsule contains no network-call stubs, socket structures, or DNS-resolution entries.

    \item No general resource-operation capabilities: the capsule does not directly expose general system calls such as $fork$, $exec$, $mmap$, $dlopen$, or $socket$, and it cannot bypass the Intent ABI to produce arbitrary external effects. However, the capsule still retains restricted local computation, inference, planning, and state-update capabilities.

    \item Mandatory DSL constraints: high-risk agents may be required to use a restricted domain-specific language (DSL) whose syntax permits only composition of intent primitives declared in the Manifest, thereby reducing intent drift at the language level.
\end{itemize}

The capsule runtime can be implemented using WASM, eBPF, native code restricted by CFI/SFI, or hardware-capability architectures such as CHERI. This paper does not elaborate on a specific engineering implementation; instead, it emphasizes that these mechanisms serve “capability-expression reduction” in AgenticOS, not merely traditional code isolation.

\subsection{Semantic Boundary Gateway}
The Semantic Boundary Gateway is the only controlled boundary between Agent Capsules and the external world. Its role is not to infer an agent’s true intent, but to transform structured intents into controlled external effects after the Logic Shutter has authorized them.

The Semantic Boundary Gateway includes the following functions:

\begin{itemize}
    \item Protocol proxying: complex protocols such as SSH, TLS, HTTP, message queues, and object storage are implemented on the gateway side, while Agent Capsules do not access protocol-stack details.

    \item Credential hosting: private keys, access tokens, certificates, and session cookies are hosted by the gateway. Capsules receive only task-bound semantic capabilities, not copyable credentials.

    \item Semantic channels: capsules and the gateway do not exchange raw byte streams. They communicate through structured messages. All fields are subject to strict schema validation, and blob fields are bound to declarative digests to avoid protocol tunneling.

    \item Output auditing and normalization: the gateway performs content filtering, structural validation, size padding, and pacing control for all outgoing content, reducing data-exfiltration risk and covert-channel bandwidth.
\end{itemize}

From the perspective of the security boundary, the Semantic Boundary Gateway turns the “external world” from a directly reachable object into a system-hosted semantic resource. An agent cannot arbitrarily construct TCP byte streams; it can only call Manifest-bound Intent ABI functions such as $FetchJSON$, $UploadArtifact$, and $SendMessage$.

%% file: chapter/intent.tex
\section{Intent ABI and Capability Generation Mechanism}

\subsection{Manifest-Only Runtime}
The success of AgenticOS depends on whether its semantic interfaces can remain “non-generalizable,” meaning that the compositional closure of the interface set must not degenerate into a general system-call table. The Manifest-Only Runtime is the core mechanism for achieving this goal.

A Manifest contains at least the following information:

\begin{itemize}
    \item Agent identity and version: agent name, model or code version, build hash, and signature information.

    \item Input objects: readable documents, tables, database views, message queues, or external APIs.

    \item Output targets: writable documents, artifact repositories, message channels, or approval systems.

    \item Network boundary: allowed domains, path prefixes, HTTP methods, request rates, and data types.

    \item Tool capabilities: permitted compilation, testing, template rendering, static analysis, or patch-application capabilities.

    \item Model capabilities: permitted model sessions, context windows, data-retention policies, and inference budgets.

    \item Information-flow rules: allow, deny, or require-human-approval relationships between input source labels and output sinks.

    \item Human-confirmation points: explicit human confirmation for high-risk external effects, such as sending emails, committing code, deploying services, or transferring funds.
\end{itemize}

A Manifest is not an ordinary configuration file; it is the sole basis for capability synthesis. Without a Manifest declaration, there is no corresponding ABI Stub. Without an ABI Stub, there is no entry point for producing that class of external effect.

\subsection{Design Principles of the Intent ABI}
The Intent ABI is an application binary interface oriented toward task intent. It differs from a traditional syscall ABI: the latter provides general resource operations, whereas the former provides structured semantic operations.

Its design principles are as follows:

\begin{itemize}
    \item Non-generalizability: interface composition must not express arbitrary I/O, arbitrary process creation, or arbitrary network protocols.

    \item No byte streams: primitives such as $socket$, $pipe$, and raw file descriptors are prohibited.

    \item No arbitrary execution: general execution such as $exec(cmdline)$ is prohibited; toolchain capabilities must be decomposed into concrete and auditable semantic interfaces.

    \item Reified capabilities: network permissions are bound to the five-tuple $\langle$\emph{domain}, \emph{path\,prefix}, \emph{method}, \emph{data\,type}, \emph{budget}$\rangle$.

    \item Auditable output: all external effects must be recorded as structured events and associated with the Manifest and capability tokens.

    \item Constrained information flow: interface inputs and outputs must carry labels, enabling the policy engine to analyze propagation paths from sources to sinks.
\end{itemize}

\subsection{Core Interface Families}

The Intent ABI can be divided into six families shown in Table~\ref{tab:intent-abi}.

\begin{table}[t]
\centering
\caption{Intent ABI families and their primitives.}
\label{tab:intent-abi}
\renewcommand{\arraystretch}{1.25}
\begin{tabularx}{\columnwidth}{@{}lX@{}}
\toprule
\textbf{Family} & \textbf{Primitives} \\
\midrule
Identity \& Session    & \texttt{GetAgentIdentity}, \texttt{GetExecutionContext}, \texttt{EmitAuditEvent} \\
Documents \& Data      & \texttt{ReadDeclaredDoc}, \texttt{WriteDeclaredDoc}, \texttt{ParseDoc}, \texttt{RenderDoc}, \texttt{TransformTable}, \texttt{ComputeHash} \\
Network Capabilities   & \texttt{FetchJSON}, \texttt{FetchDoc}, \texttt{UploadArtifact}, \texttt{SendMessage}, \texttt{ResolveDNS}, \texttt{RequestHumanApproval} \\
Toolchain              & \texttt{PythonCompile}, \texttt{RunUnitTest}, \texttt{RenderTemplate}, \texttt{ApplyPatch}, \texttt{BuildArtifact}, \texttt{StaticAnalyze} \\
Model Inference        & \texttt{OpenModelSession}, \texttt{ModelInfer}, \texttt{ModelEmbed}, \texttt{ModelSummarize}, \texttt{CloseModelSession} \\
Environment Observation & \texttt{GetResourceBudget}, \texttt{GetTrustedTime} \\
\bottomrule
\end{tabularx}
\end{table}

Each interface call must carry the $cap\_id$ declared in the Manifest. The policy engine then performs fine-grained validation and makes a mediation decision based on information-flow labels, resource budgets, call history, and human-confirmation state.

\subsection{Prohibited Interfaces}

The following POSIX semantics are permanently excluded from the Intent ABI: $open$, $read$, $write$, $socket$, $connect$, $bind$, $mmap$, $execve$, $fork$, $ptrace$, $dlopen$, and $syscall$.

This list is not merely an API blacklist; it is the lower bound of the security boundary. It provides hard termination rules for static analyzers, the Weaver linker, and formal verification: any component that attempts to inject raw file descriptors, raw sockets, dynamic linkers, or arbitrary execution entries into an Agent Capsule must fail during the build stage.

\subsection{Weaver and Dynamic ABI Generation}
Weaver is the builder in AgenticOS that generates the capsule runtime environment from the Manifest, agent code, and standard semantic library. Its role is not merely to link libraries, but to prove that “the capability set an agent can call” is consistent with “the task boundary declared by the Manifest.”

\begin{figure*}[h]
    \centering
    \includegraphics[width=1.0\linewidth]{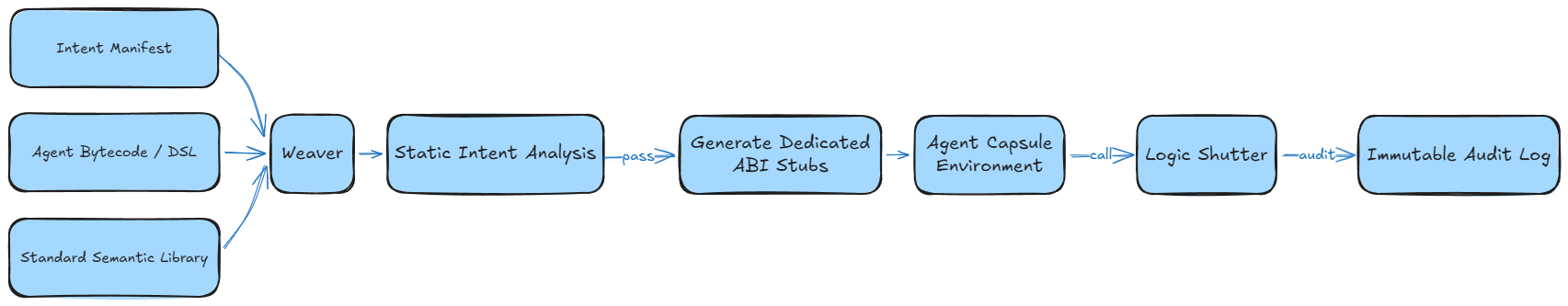}
    \caption{Generation and Execution Workflow from Manifest to Dynamic ABI}
    \label{fig:Workflow}
\end{figure*}

Weaver must perform the following checks:

\begin{itemize}
    \item Consistency between Manifest and code references: the code must contain no call sites for undeclared capabilities.

    \item Valid standard semantic-library signatures: all ABI Stubs must have verifiable origins and traceable versions.

    \item Unreachability of prohibited interfaces: the capsule binary must contain no raw syscalls, dynamic loaders, raw sockets, or external interpreter entries.

    \item Analyzable information-flow paths: paths from input labels to output sinks must comply with organizational policies.

    \item Reproducible build: external auditors must be able to reproduce the capsule environment and verify that the ABI Stubs are fully consistent with the Manifest.
\end{itemize}

\subsection{Skill Generation, Registration, and Admission}
AgenticOS allows agents to assist in generating new Skills in the development state. A Skill here refers to an operating-system-native capability unit callable through the Intent ABI, such as “organize a table and generate a report,” “run project unit tests,” “perform a read-only query against a specified API,” or “generate an email draft from a template.”

However, Skill generation must not be equivalent to runtime self-escalation. Before any new Skill enters the AgenticOS runtime state, it must pass the following admission process:

\begin{itemize}
    \item Development-state generation: an agent may generate Skill code, a draft Manifest, and test cases in an uncontrolled or low-trust development environment.

    \item Capability declaration: the Skill must declare its required inputs, outputs, network boundary, toolchain dependencies, model capabilities, and human-confirmation points.

    \item Static analysis: Weaver checks whether the Skill references prohibited interfaces, contains undeclared external effects, or violates information-flow policies.

    \item Testing and proof: the Skill must provide unit tests, behavioral examples, code-intent consistency proofs, or equivalent audit materials.

    \item Signature and registration: an approved Skill is signed and registered as a component of the system semantic library.

    \item Runtime invocation: agents can invoke registered Skills only through Manifest declarations and capability tokens.
\end{itemize}

Thus, the essence of the AgentOS-native Skill ecosystem is not to let agents expand capabilities without limits, but to let application capabilities settle into system capabilities in an auditable, composable, and revocable manner.

%% file: chapter/security.tex
\section{Security Analysis}
\label{execution}

\subsection{Structural Removal of the System-Call Attack Surface}
Traditional container escapes and kernel privilege escalations usually rely on exploitable system-call interfaces, complex kernel objects, or device-driver flaws. In AgenticOS, the Agent Capsule address space contains no general system-call entry, no direct page-table access capability, no raw protocol stack, and no dynamic linker. The Ghost Kernel itself also exposes no runtime service interface to capsules.

Therefore, under the stated security assumptions, traditional syscall-based escape paths are structurally removed. More precisely, escape is no longer expressed as “calling a dangerous kernel interface from inside a capsule,” but is transformed into the following narrower questions:

\begin{itemize}
    \item whether the Ghost Kernel isolation implementation has defects;

    \item whether the Logic Shutter has incorrectly mediated an intent;

    \item whether Weaver has generated ABI Stubs inconsistent with the Manifest;

    \item whether the Semantic Boundary Gateway has incorrectly exposed raw protocols or credentials.
\end{itemize}

Residual risks remain. For example, if the Ghost Kernel contains page-table management errors, DMA-control flaws, or measurement-chain vulnerabilities, it may still become an attack entry. Therefore, the Ghost Kernel should adopt minimization and formal-verification ideas similar to seL4, and combine them with hardware TEEs, IOMMU, and memory encryption to reduce the impact of implementation flaws.

\subsection{Supply-Chain Poisoning and Capability-Composition Attacks}
Even if each individual Intent ABI appears safe, legitimate combinations of multiple ABIs may still construct unintended malicious behavior. For example:

\lstdefinestyle{callchain}{
  basicstyle=\ttfamily\small,
  backgroundcolor=\color{gray!5},
  frame=single,
  xleftmargin=1em,
  columns=fullflexible,
  keepspaces=true,
  moredelim=[is][\color{blue!70!black}\bfseries]{>>}{<<},
}
\begin{lstlisting}[style=callchain]
FetchJSON
  -> ModelSummarize
  -> WriteDeclaredDoc
  -> UploadArtifact | SendMessage
\end{lstlisting}

Such an attack may appear as a normal call sequence at the level of single-call auditing. AgenticOS therefore cannot merely check whether each individual interface is authorized; it must also analyze data flow and the intent-interaction graph between interfaces.

Hardening measures include:

\begin{itemize}
    \item Intent-interaction graph analysis: during static analysis, construct the agent’s intent-interaction graph and identify paths that may form “sensitive input -- model transformation -- external output” chains.

    \item Information-flow labeling: attach source labels to data obtained through interfaces such as $FetchJSON$ and $ReadDeclaredDoc$, and enforce label policies at output interfaces such as $UploadArtifact$ and $SendMessage$.

    \item Human approval: require $RequestHumanApproval$ for high-risk flow paths and write the approval result into the audit chain.

    \item Runtime behavior baselines: the Logic Shutter maintains baselines for call frequency, data-flow direction, output-size distribution, and tool-composition patterns; deviations trigger blocking or degradation.
\end{itemize}

\subsection{Intent Drift}
Intent drift refers to a mismatch between the Manifest submitted by an agent and the agent’s actual behavioral objective. For example, an agent may claim to perform “data analysis” while its actual code includes network scanning, credential probing, or covert exfiltration logic. Because predicting arbitrary code behavior is undecidable, static analysis cannot completely rule out such drift in finite time.

AgenticOS reduces this risk in layers:

\begin{itemize}
    \item High-risk agents are required to use a restricted DSL, so that the language itself contains only the intent primitives declared in the Manifest.

    \item Agents written in general-purpose languages are subject to conservative static analysis, and all unrecognized calls are rejected by default.

    \item Agents that pass admission are subject to stricter runtime monitoring, resource budgets, and output auditing.

    \item Interfaces involving external side effects require mandatory human confirmation or policy approval.
\end{itemize}

This means that AgenticOS does not claim to fully “understand” the true intent of arbitrary agents. Instead, through language constraints, capability generation, and runtime mediation, it makes undeclared intent difficult to convert into external effects.

\subsection{Covert Channels}
Although AgenticOS prohibits raw byte streams and protocol tunneling, covert channels cannot be completely eliminated. Residual channels include:

\begin{itemize}
    \item Timing channels: an agent encodes information by controlling the timing interval between API calls.

    \item Size channels: an agent transmits bits by choosing different lengths of otherwise legitimate outputs.

    \item Presence channels: the choice of whether to call an interface can itself encode information.

    \item Content-selection channels: an agent selects among multiple legitimate summaries, phrasings, or orderings to transmit information at low bandwidth.
\end{itemize}

Mitigations include:

\begin{itemize}
    \item Request-timing normalization: the Logic Shutter batches external requests into fixed time windows.

    \item Output-size padding: the Semantic Boundary Gateway transmits structured outputs in fixed-size blocks and pads insufficient blocks with randomness.

    \item Output-template constraints: in high-sensitivity scenarios, outputs must conform to fixed schemas, enumerated fields, or templates.

    \item Channel-bandwidth limitation: rate limiting, budget management, and output normalization reduce residual channel bandwidth to a level that is difficult to exploit in practice.
\end{itemize}

Thus, the goal of AgenticOS with respect to covert channels is not to eliminate them completely in theory, but to compress them from high-bandwidth, hard-to-audit general communication channels into low-bandwidth, modelable, and monitorable residual risks.

\subsection{Trusted Boundaries of the Logic Shutter and Weaver}
The Logic Shutter centrally hosts intent parsing, policy enforcement, capability tokens, information-flow labels, and audit logic. Its TCB complexity is higher than that of the Ghost Kernel. Weaver, meanwhile, extracts interfaces from the standard semantic library and injects them into capsules. If the standard semantic library is polluted—for example, if it contains an undocumented $RawFD$ interface—the generated ABI may accidentally open a backdoor for agents.

Hardening measures include:

\begin{itemize}
    \item TCB minimization by decomposition: split the Logic Shutter into multiple isolated components, such as a DNS shutter, HTTP shutter, file shutter, model shutter, and policy core, each running in an independent isolation domain.

    \item Implementation in memory-safe languages: implement the core Logic Shutter and Weaver in memory-safe languages such as Rust to reduce buffer-overflow and use-after-free risks.

    \item Standard-library signatures and reproducible builds: all semantic-library components are cryptographically signed, and Weaver must verify the signature chain before injection.

    \item Manifest-code binding proof: agent submissions include a code-intent consistency proof, which Weaver machine-verifies during admission.

    \item Formal verification of critical paths: formally model the policy engine, capability-token manager, and ABI prohibited-interface checks.
\end{itemize}

%% file: chapter/application.tex
\section{Application Capability Migration and Ecological Boundaries}

\subsection{From Application Migration to Application-Capability Migration}

Traditional operating systems mainly provide low-level abstractions such as files, processes, networks, and devices to applications. Traditional applications then encapsulate business logic, interaction interfaces, and external-service connections on top of those abstractions. The design innovation of AgenticOS lies in changing the boundary between the operating system and application capabilities: it attempts to extract from individual applications those software capabilities that can be declared, audited, and constrained, and consolidate them into governable semantic capabilities on the system side.

Therefore, the migration target is not the “application itself,” but the “application capability unit.” An application often plays three roles at once: human-computer interface, business-capability carrier, and resource-access proxy. AgenticOS does not require these three roles to migrate together. Instead, it prioritizes capability units that can be described by a Manifest, invoked through the Intent ABI, bound to capability tokens, and recorded as audit events. In this way, the operating system is no longer merely a resource provider for applications, but gradually becomes the hosting layer, governance layer, and composition layer for application capabilities.

The key value of this migration approach is that application capabilities move from “private functions inside an app” to “system-side capability objects that are declarable, reviewable, revocable, and composable.” It preserves necessary human-computer interfaces and specialized tool forms, while preventing agents from needing to simulate UIs, abuse scripts, or directly touch low-level system calls to complete tasks. In the long run, the goal of AgenticOS is not to cover all application experiences, but to cover an increasing number of software capabilities that are delegable, amenable to semantic abstraction, and auditable.

\subsection{AgentOS-Native Skill Ecosystem}

As more private application capabilities settle into AgentOS-native Skills, the boundary of operating-system capabilities also changes. Traditional operating systems provide abstractions for files, processes, networks, and devices. AgenticOS further provides high-level semantic capabilities such as “organize documents,” “run tests,” “generate reports,” “query specified data sources,” and “send approval requests.”

However, the Skill ecosystem must follow two principles:

\begin{itemize}
    \item Skills may grow, but the runtime capabilities of an agent may not grow spontaneously. New Skills must pass development-state generation, review, signature, and registration processes.

    \item A Skill is a system capability, not an application-private backdoor. Every Skill must expose its Manifest, input/output schema, information-flow rules, audit events, and revocation mechanism.
\end{itemize}

Under this model, AgenticOS can gradually absorb delegable software capabilities from traditional applications while avoiding placing agents back on top of infinitely composable general interfaces.

%% file: chapter/discussion.tex
\section{Discussion and Limitations}
\label{Discussion and Limitations}

\subsection{Completeness of Intent Expression}
The true intent of arbitrary code cannot be precisely determined in finite time; this is a variant of the halting problem and program-semantics undecidability. AgenticOS therefore cannot rely on “automatically understanding all code intent” as its security foundation. Feasible paths include enforcing restricted DSLs, deriving conservative approximations through static analysis, sampling agent behavior in the development state to assist Manifest generation, and limiting external effects at runtime through capability tokens and audit mechanisms.

\subsection{Performance and Usability Trade-Offs}
Each semantic call may trigger multiple context switches from the Agent Capsule to the Logic Shutter, the Semantic Boundary Gateway, and the Ghost Kernel, incurring higher overhead than traditional system calls. Batched asynchronous requests, ABI caching, policy precompilation, and fixed-time-window processing can mitigate performance issues, but may also increase latency. AgenticOS must balance security, usability, and interaction experience.

\subsection{POSIX Ecosystem Migration Cost}
Abandoning POSIX compatibility means that traditional software cannot run unmodified inside Agent Capsules. This paper does not advocate migrating the entire software ecosystem at once. Instead, it advocates prioritizing software capabilities that are delegable, amenable to semantic abstraction, and auditable. Existing applications can continue to serve as development-state tools, human-computer interfaces, or sources of capabilities; verified capabilities can then be consolidated into the Intent ABI or AgentOS-native Skills.

\subsection{Boundaries of Formal Verification}
The current architecture requires formal verification of the Ghost Kernel, critical paths in the Logic Shutter, and Weaver’s prohibited-interface checks. However, complete end-to-end verification from Manifest declarations to actual capsule behavior remains difficult. Future work includes establishing composition-safety theorems for the Intent ABI, proving information-flow constraints, and building a verifiable supply chain from Skill registration to runtime invocation.

\subsection{Human Confirmation, Authorization Closure, and Responsibility Boundaries}

AgenticOS addresses the boundary of agent capabilities, not responsibility attribution in the real world. Even if the system can restrict an agent’s external effects through the Manifest, capability tokens, and the Logic Shutter, high-risk decisions still require a clearly identifiable authorizing actor and responsibility anchor. In other words, AgenticOS can answer the question “whether this agent is allowed to initiate a certain class of operation,” but it cannot transfer final responsibility for contract signing, fund transfers, production changes, or the adoption of medical advice to an automated executor.

Therefore, human confirmation should not be treated as an ad hoc patch at the interaction layer, but as a first-class capability in the Intent ABI. Interfaces such as $RequestHumanApproval$ serve at least three functions: first, they explicitly interrupt high-risk external effects in the agent’s automatic execution chain; second, they allow a user or organizational owner to confirm whether the current operation matches the true intent, risk preference, and business context; third, they write the confirmation result, confirmer, confirmation scope, and policy version into the audit chain, forming a traceable authorization closure.

This design does not weaken the autonomy goal of AgenticOS. On the contrary, it defines the system boundary between autonomous agents and human responsibility: low-risk, reversible, and auditable capabilities may be executed automatically by agents; high-risk, irreversible, or real-world commitment-bearing capabilities must enter the system through explicit confirmation. Human confirmation is therefore not an exception mechanism in AgenticOS, but a key interface in its security model that connects technical control with real-world responsibility.

%% file: chapter/conclusion.tex
\section{Conclusions}

This paper proposes AgenticOS, a conceptual architecture that reconstructs operating-system security boundaries at a fundamental level. Its core contribution is not to propose a stronger sandbox, but to migrate the operational boundary of agents from resource primitives to intent primitives. Agents no longer directly obtain raw file, raw network, raw process, or raw protocol capabilities. Instead, they produce controlled external effects through the Manifest, Intent ABI, capability tokens, Logic Shutter, and Semantic Boundary Gateway.

In this model, the Ghost Kernel provides the minimal trusted isolation substrate; the Logic Shutter performs intent mediation and information-flow control; the Agent Capsule hosts the restricted runtime; the Semantic Boundary Gateway proxies external protocols and credentials; and Weaver ensures consistency between the Manifest and capsule capabilities. AgenticOS does not promise to eliminate all risks, nor does it attempt to replace all application forms. It aims to provide a more appropriate operating-system security abstraction for the era of autonomous agents: enabling software capabilities that are delegable, amenable to semantic abstraction, and auditable to gradually become operating-system-native capabilities, while preserving necessary human-computer interfaces, specialized tool forms, human-confirmation mechanisms, and responsibility boundaries.